\documentclass[sigconf]{acmart}
\usepackage{booktabs}
\usepackage{array}
\newcolumntype{M}[1]{>{\centering\arraybackslash}m{#1}}
\usepackage{tipa}
\usepackage{ragged2e}
\AtBeginDocument{%
  \providecommand\BibTeX{{%
    \normalfont B\kern-0.5em{\scshape i\kern-0.25em b}\kern-0.8em\TeX}}}

\copyrightyear{2024}
\acmYear{2024}
\setcopyright{rightsretained}
\acmConference[CHI '24]{Proceedings of the CHI Conference on Human Factors in Computing Systems}{May 11--16, 2024}{Honolulu, HI, USA}
\acmBooktitle{Proceedings of the CHI Conference on Human Factors in Computing Systems (CHI '24), May 11--16, 2024, Honolulu, HI, USA}
\acmDOI{10.1145/3613904.3642775}
\acmISBN{979-8-4007-0330-0/24/05}

\begin{document}
\RaggedRight

\title{From Fitting Participation to Forging Relationships: The Art of Participatory ML}

\author{Ned Cooper}
\email{edward.cooper@anu.edu.au}
\orcid{0000-0003-1834-279X}
\affiliation{%
  \institution{Australian National University}
  \city{Canberra}
  \country{Australia}
}

\author{Alexandra Zafiroglu}
\email{alex.zafiroglu@anu.edu.au}
\orcid{0000-0003-0918-3977}
\affiliation{%
  \institution{Australian National University}
  \city{Canberra}
  \country{Australia}
  }

\renewcommand{\shortauthors}{Cooper and Zafiroglu}

\begin{abstract}
\justifying
Participatory machine learning (ML) encourages the inclusion of end users and people affected by ML systems in design and development processes. We interviewed 18 {\itshape participation brokers}---individuals who facilitate such inclusion and transform the products of participants' labour into inputs for an ML artefact or system---across a range of organisational settings and project locations. Our findings demonstrate the inherent challenges of integrating messy contextual information generated through participation with the structured data formats required by ML workflows and the uneven power dynamics in project contexts. We advocate for evolution in the role of brokers to more equitably balance value generated in Participatory ML projects for design and development teams with value created for participants. To move beyond `fitting' participation to existing processes and empower participants to envision alternative futures through ML, brokers must become educators and advocates for end users, while attending to frustration and dissent from indirect stakeholders.
\end{abstract}

\maketitle

\justifying

\section{Introduction}
A growing body of research and practice in the field of participatory machine learning (`Participatory ML') encourages the inclusion of end users and people affected by ML systems in design and development processes \cite{Kulynych2020-et}. This nascent field builds on a long history of participatory approaches to computing research and development and has emerged in response to examples of sub-par performance of ML systems for marginalised groups. Participatory approaches have been enacted across each stage of ML design and development---from problem formulation to model evaluation---and include collaborative approaches to construct datasets \cite{Theodorou2021-sx,Pushkarna2022-pm}, design and validate ML algorithms \cite{Suresh2022-vv,Nekoto2020-rv}, and guide advocacy for algorithmic accountability \cite{Queerinai2023-py,Katell2020-ph}. At the same time, several authors have raised concerns about ``participation-washing'' \cite{Sloane2022-ev}, co-optation of participatory work \cite{Birhane2022-hx}, and the limited evidence across Participatory ML projects of equitable partnerships with participants \cite{Delgado2023-tm,Feffer2023-jx,Corbett2023-de}.

In this paper, we take a ``studying up'' lens \cite{Nader1972-cb}, focusing on the researchers, practitioners, and activists who facilitate the inclusion of participants in the development of an ML artefact or system and implement feedback from those participants into development processes. We describe these individuals as {\itshape participation brokers}, and the work they lead and champion as {\itshape participatory projects}. Such projects sit within the context of new or ongoing ML development and the work of brokers is bifocal; deciding and facilitating who becomes a participant and what they participate in, and transforming the products of participants' labour into inputs for an ML artefact or system.

Despite growing research in the field, a gap exists in the literature regarding how brokers facilitate meaningful participation and manage power dynamics in successful Participatory ML projects. This limits our understanding of the practical strategies and challenges involved in implementing effective projects. To address this gap, we conducted 18 interviews with participation brokers from a range of organisational settings and project locations to understand the different incentives, strategies, challenges and outcomes of participatory work in the varied contexts of ML development. In particular, we committed to expanding the scope of Participatory ML by including projects in our study based outside North America and Europe, which have been most reported to date.

Specifically, three exploratory questions guided our interviews:
\begin{enumerate}
    \item Who is defined as a participant and included in Participatory ML projects?
    \item In what stages of the ML design and development process are those participants included?
    \item How is feedback from participants during those stages implemented?
\end{enumerate}

Through a reflexive thematic analysis, we demonstrate the challenges brokers encounter in `fitting' participants into ML workflows, and we describe tensions between the messy nature of participatory work and the rigid structure of ML data formats. We conclude with a reflection on our findings, advocating for evolution in the role of brokers to more equitably balance value generated in Participatory ML projects for design and development teams with value created for participants. We argue that projects that explicitly include education for and engagement with end users, and create processes for mediating frustration and dissent of indirect stakeholders within the limits of the ML development process, demonstrate how brokers can move beyond fitting outside voices into established engineering processes.

\section{Background} \label{Background}

\subsection{Participation in Computing Research and Development}

The participation of intended users and stakeholders affected by a policy, product or intervention has long been a principle pursued across diverse fields, such as international development and environmental justice. Participatory approaches have also been a feature of computing research and development for many decades, with roots in Action Research (AR) \cite{Hayes2011-wk} and the rise of Participatory Design (PD) in Scandinavia in the 1970s \cite{Sundblad2011-td,Schuler1993-wu}. PD was initially driven by the desire of labour unions to have input into how computer systems were introduced in the workplace \cite{Kensing1998-xn,Gregory2003-yq}. The participatory approach was not only a methodological choice but also a political stance, advocating for the democratisation of design and acknowledging the value of domain expertise \cite{Bjorgvinsson2010-sg}.

In HCI, PD developed into a distinct method for merging the theoretical knowledge of researchers and designers with the practical knowledge of users \cite{Muller2012-jz,Muller1993-eb,Bodker2006-ej}, while Value Sensitive Design broadened HCI's scope to include not only direct users but also indirect stakeholders \cite{Friedman1996-ub,Friedman2019-cn}. Over recent decades, there has also been a shift in HCI towards community-collaborative approaches to research, expanding beyond engagement with individual users and towards engagement with groups and communities \cite{Cooper2022-fh,DiSalvo2013-ut}.

\subsection{The Emergence of Participatory ML}

Academics and journalists have documented algorithmic bias and discrimination for decades, as part of critical internet studies, though discourse in this area has accelerated since the `techlash' in 2018 \cite{Dubber2020-nc}. Over the past five years, this critical discourse has uncovered many examples of sub-par performance of ML systems for marginalised groups, including women \cite{Buolamwini2018-yl,Bolukbasi2016-vy,Savoldi2021-nl}, people of colour \cite{Noble2018-ic,Benjamin2019-ou}, people with a disability \cite{Hutchinson2020-mx,Venkit2022-uw}, and LGBTQ+ communities \cite{Keyes2018-tz}.

The growing field of AI ethics has been criticised, however, for focusing inwardly on technologists and their agency \cite{Greene2019-qy}. Research in AI ethics is often limited to lab environments, where technologists design, develop, and, perhaps most importantly, test systems, with ``the intention of deploying those systems in a different context'' \cite{Gansky2022-cq}. \citet{Hoffmann2019-uv} argues that such work is subject to the same limitations as anti-discrimination law in addressing racial discrimination, which arises from deeper inequalities in social systems. Similarly, \citet{Selbst2019-il} note that the abstraction of problems from social contexts to develop a computing response fails to consider how social contexts affect the distribution and use of computing systems.

The turn to participatory approaches to developing ML systems may be understood as part of a broad shift away from the inward focus of some AI ethics discourse, and towards understanding and addressing the lived experience of data subjects\footnote{Data subjects are defined in this paper as people whose personal data can be used to identify them. Data subjects include end users, defined in this paper as the direct users of an ML artefact or system. However, a person does not need to be a direct user of a data-driven system to be a data subject. Many data subjects are tracked, scored, and analysed indirectly by data-driven systems---{\itshape e.g.}, subjects of algorithmic credit scoring systems \cite{Ziewitz2021-sl}}. Proponents of Participatory ML encourage a shift in research and development activities out of labs and into the communities in which people live, to build more ``inclusive'' and ``democratic'' systems \cite{Kulynych2020-et}. \citet{Katell2020-ph} argue that the shift is driven in part by end users, who are starting to demand control over algorithms embedded in products and services they interact with as they have previously demanded control over their data---control over what models optimise for and an understanding of how the models work.

The growth of Participatory ML is exemplified by its inclusion in papers [{\itshape e.g.}, \citealp{Lee2019-ke,Halfaker2020-jt,Bondi2021-fz}], panels \cite{Zytko2022-uy}, and workshops at computing conferences---most significantly, the Participatory ML workshop at the 2020 International Conference on Machine Learning (ICML) \cite{Kulynych2020-et}. Additionally, industry-based research teams are considering how to enable Participatory ML, by broadening the circle of people who can engage in the development of ML systems [{\itshape e.g.}, \citealp{Google_People_AI_Research_PAIR2020-iu}]. Participation has also emerged as a key principle for government and civil society stakeholders in ML-related research, development, and monitoring [{\itshape e.g.}, \citealp{Farthing2021-hr,Hu2019-rc,White_House_Office_of_Science_and_Technology_Policy2021-bb,Berditchevskaia2021-lh}].

\subsection{Power and Participatory ML}

The turn to Participatory ML has been accompanied, however,  by cautions against ``participation-washing'' that have beset other fields of design \cite{Sloane2022-ev}. \citet{Sloane2022-ev} argue that the ``ML community'' must interrogate all forms of inclusion in ML design and development, including those that are passive or extractive---where people become data points in datasets used to train ML systems, with little or no reciprocal value \cite{Mayer-Schonberger2013-yz}. Similarly, \citet{Birhane2022-hx} retrace a genealogy of participation to highlight potential limitations of and risks for Participatory ML, such as the conflation of inclusion and participation and the co-optation of participatory work by corporate actors.

Since the ICML workshop, several authors have reviewed Participatory ML projects and demonstrated a gap between the field's goals and actual practices. \citet{Feffer2023-jx} and \citet{Robertson2020-li} argue that many Participatory ML projects are limited to developing computational methods to elicit participant preferences, rather than re-imagining participants as co-designers throughout the ML process. Similarly, \citet{Corbett2023-de} and \citet{Delgado2021-pg} draw on Arnstein's Ladder of Citizen Participation \cite{Arnstein1969-rc} to note the lack of consistency in what gets classified as a participatory approach in ML development, and a tendency  towards informing or consulting with participants, rather than engaging in equitable partnerships. Others have raised specific concerns about projects in industry settings, including the use of algorithmic proxies for participants \cite{Delgado2023-tm} and a lack of transparency about practices or adequate incentives for sharing power with participants \cite{Groves2023-pe}.

Some authors have proposed frameworks or guiding questions for Participatory ML, to shift practice towards more equitable partnerships with participants \cite{Delgado2023-tm, Feffer2023-jx, Corbett2023-de}, and to ensure participation informs institutional decisions \cite{Zhang2023-hz}. These frameworks address concerns that participation alone cannot redistribute power and, without attention, may be more instrumental for researchers than beneficial for participants, reflecting longstanding discussions in PD about other computing systems \cite{Muller1993-eb,Vines2013-bt,Bratteteig2014-co}.

While existing research has investigated the risks and limitations of Participatory ML, and highlighted a gap between principles and practice, successful case studies of meaningful participation have nevertheless gained recognition within the research community. This paper seeks to answer: what are the incentives, strategies, and challenges of brokers in facilitating such meaningful forms of participation? If participatory approaches are fundamentally about dismantling power and information asymmetries between data scientists and end users (or other stakeholders), how do participation brokers actively manage power dynamics throughout the ML development process? Our research contributes an outline of how brokers exercise and manage their power through their decisions and actions in projects recognised by the research community for their collaborative approach. We hope these contributions provide guidance for future participation brokers across diverse ML development processes, in varied organisational settings and project locations.

\section{Method}

We used expert interviews to derive interpretations of participatory work, rather than facts, geared toward meaning-making by the researchers \cite{Warren2001-gz}. In-depth interviews provide ``thick'' descriptions \cite{Geertz1973-ao} of the experience of conducting participatory work, beyond what is reported in publications. Below we outline our approach to recruitment, data collection and data analysis. We end this section by reflecting on our positionality in relation to the study and outlining the limitations of our methodological approach.

\subsection{Recruitment}

To recruit interviewees, we used a two-step approach combining purposeful and snowball sampling \cite{Palinkas2015-tn}. First, we contacted authors of papers published at the aforementioned ICML workshop that reported on information, datasets, or models developed from Participatory ML projects. We then asked these authors to recommend other relevant projects for our consideration. Second, we contacted authors of projects identified by our first-round interviewees, prioritising projects conducted outside North America and Europe. We also asked for recommendations for additional projects from our second-round interviewees.

\subsection{Data Collection}
We conducted 18 semi-structured, 60-minute interviews via video conference in November 2022 and the first half of 2023. All interviews were approved by a Human Research Ethics Committee, transcribed, and shared with interviewees for review. Table 1\footnote{Four projects traversed two locations. We document the location to which our interview primarily related; NFP = Not-for-profit; Several projects included multiple artefacts comprising a system, rather than one artefact. We document those projects in the `model' category.} provides details about interviewees and related projects, with some information removed to preserve anonymity.

While each interview followed the same general format, covering descriptive and reflective questions, we tailored the interview guides based on each interviewee's background and published work. Descriptive questions focused on the stages and people involved in the interviewee's project(s), while reflective questions asked about the interviewee's incentives for, challenges in, and outcomes of participatory practices. The interviewer also facilitated open-ended discussion by probing interviewees for details and reflections when interviewees mentioned examples of their decisions and actions in projects.

\begin{table}[ht]
\centering
\begin{tabular}{cM{1.8cm}M{3.2cm}M{1.6cm}}
\toprule
\textbf{ID} & \textbf{Project Location} & \textbf{Organisational Setting} & \textbf{Key Artefact} \\
\midrule
P1 & USA & Industry & Model \\
P2 & Oceania & Academia & Information \\
P3 & USA & Civil society (NFP) & Information \\
P4 & Latin America & Academia & Model \\
P5 & South Asia & Civil society (NFP) & Information \\
P6 & Africa & Government & Model \\
P7 & Europe & Academia & Dataset \\
P8 & USA & Civil society (Advocacy) & Information \\
P9 & Africa & Academia & Model \\
P10 & USA & Academia & Model \\
P11 & Europe & Academia & Information \\
P12 & USA & Academia & Information \\
P13 & USA & Academia & Model \\
P14 & South Asia & Civil society (NFP) & Model \\
P15 & USA & Industry & Dataset \\
P16 & USA & Academia & Dataset \\
P17 & USA & Academia & Model \\
P18 & South Asia & Civil society (Advocacy) & Information \\
\bottomrule
\end{tabular}
\caption{Interviewee details}
\label{tab:interviewees}
\end{table}

\subsection{Data Analysis}
Reflexive thematic analysis encourages researchers to embrace their subjectivity as they interpret themes from the data \cite{Braun2006-nn}. Coding in reflexive thematic analysis is an active and reflexive process that bears the mark of the researcher(s) \cite{Braun2019-me}. We used an experiential and inductive approach, focusing on both semantic and latent content to generate themes from the analysis \cite{Byrne2022-gv,Braun2020-xw}

Coding was flexible, evolving throughout the coding process. After conducting the interviews, the first author reviewed all transcripts and openly developed codes based on the semantic content of the transcripts and latent content of the interviews from their notes (initially by hand, in free-form, and later using coding software for subsequent rounds of review). During code development, the first author met in multiple rounds with the other author to share and collaboratively revise the codes. Following three coding rounds, the authors convened to discuss relationships between the codes and generate the themes presented in this paper.

\subsection{Positionality}
Both authors have experience facilitating the inclusion of participants in processes of development within and outside computing contexts ({\itshape e.g.}, in law and policy reform processes). We are motivated by the agenda of Participatory ML, and this affects our position in relation to the interviewees---we seek to learn and build on their experiences, at the same time recognising that participation is `made' in the micro-level activities of brokers and participants in projects. Our position also influences the thematic analysis---we acknowledge our primary role in participatory projects as researchers, not participants, which grants us certain societal privileges that may not be accessible to many of the participants in the projects we studied. Finally, we note that we are based outside North America and Europe, which encouraged us to expand the geographic scope of projects included in this study.

\subsection{Limitations}
Our study has several limitations. Firstly, our study prioritised interpreting themes across projects, limiting in-depth analysis for individual projects. Secondly, we engaged directly with brokers---researchers, data scientists, and activists---and not with participants in ML projects. While this aligns with our goal of understanding the processes and challenges faced by participation brokers, it limits our insights into participants' experiences to the interpretation and reporting of brokers. Thirdly, while our sampling approach aimed to identify `successful' projects, it does not guarantee that the collaborative approaches represented were effective or equitable for all brokers or participants. Additionally, the practices and experiences represented by our sample reflect an engaged subset of participation brokers, which limits perspectives in our study of those sceptical of or new to participatory approaches in ML. Fourthly, our study includes a small sample of projects from industry and government. Despite contacting 34 potential interviewees, many from industry and government were unavailable due to workload and anonymity concerns. Lastly, despite our efforts, our study includes a small number of projects located in the Global South. We hope this study broadens Participatory ML literature to consider activities and experiences outside the Global North, which are underrepresented and may be described differently to current terms of the field.

\section{Findings}

Our study demonstrates that participatory practices are not a uniform set of activities, but rather a means used by brokers to create, structure, and manage relationships between  categories of people during the development of ML artefacts and systems. In the sections below, we highlight the themes and tensions in the work of participation brokers, illustrating how participatory projects, as both methodological and political practices, challenge and are simultaneously challenged by the established processes of ML design and development.

\subsection{Becoming Brokers, Making Participants}

All brokers we interviewed sat between ML systems and end users of or people affected by those systems. In some cases, brokers were data scientists themselves, while others worked with data scientists during the project. In civil society settings, brokers often connected affected participants with decision-makers about systems or related policies.

While interviewees acknowledged pragmatic reasons for their organisations to undertake a participatory project---such as improving the performance or legitimising the deployment of an ML artefact or system---most interviewees also expressed strong personal motivations driving their project. Several interviewees spoke of using their position to {\itshape ``open up space for others to do something with ML'' [P1]}: not only voicing concerns about a model, or deciding what it should learn, but also who should use it and to what purpose it might be applied. For example, some interviewees from the Global South spoke of a desire to {\itshape ``shape the futures of machine learning and AI in [our communities] ... we don't just want to become consumers of these technologies". [P9]}

Across the interviews it was clear that brokers `make' certain people and groups into participants. As \citet{Cooper2022-bk} notes, each stage of the ML design and development pipeline implicates and affects many actors---which of those actors should formally participate? Brokers we interviewed typically prioritised the intended users of a system, such as medical professionals using diagnostic tools. In some cases, participants were the intended beneficiaries of an ML system, rather than the direct user of the artefact produced from the project ({\itshape e.g.}, a dataset of objects for use by computer vision engineers). Notably, brokers across all settings and locations relied heavily on prior relationships with organisations or communities to recruit participants---{\itshape ``it was crucial for us to cultivate a relationship with a trusted partner, a convener." [P3]} Some brokers also mentioned leveraging established relationships with groups or organisations, instead of forming new groups for projects, as a strategy to balance power dynamics with participants.

In our interviews, brokers frequently reported difficulty engaging with indirect stakeholders \cite{Friedman2019-cn} (or `people affected by a system') relative to end users. For example, patients affected by a diagnostic tool's predictions may be considered indirect stakeholders, yet one interviewee noted that: {\itshape "We're nowhere near having patients participate in algorithm development. We're at the point now where I just hope there's more transparency." [P17]} Indifference among indirect stakeholders also complicated recruitment: {\itshape ``Recruiting people is hard… who cares enough to be part of it in the first place?'' [P11]} Some interviewees questioned the value of including indirect stakeholders, asking whether {\itshape "someone who only consumes a product needs to be included" [P13]} or highlighting the potential tension between inclusion and innovation: {\itshape ``I can't emphasise enough that innovation is not a consensus-building process." [P17]} The range of perspectives underscore the debate regarding the feasibility and desirability of incorporating voices beyond direct users of ML systems into Participatory ML projects.

\subsection{Problem Formulation as Problematising ML}

Across all settings and projects, the projects included in this study were predominately initiated by the interviewees acting as brokers. A common challenge noted by interviewees was translating general discussions with partners about organisational or community problems and interests into potential ML capabilities: {\itshape ``if you go in and say, hey, let's have a conversation about data or ... automated decision-making, people are like `what is that and why is it relevant to me?''' [P2]} Given the novelty and technical dimensions of ML, several interviewees encouraged demo-ing models to encourage people engaged as participants to imagine applications, rather than formulating ideas in the abstract: {\itshape ``Very few people understand what we're even trying to achieve until they see a model do something in practice.'' [P7]} In some cases, brokers combined prototypes with processes that engaged participants in critical questioning about the societal impact of ML systems. For example, one interviewee designed prototypes reflecting existing public ML systems to {\itshape ``probe the environment and allow us [researchers] to ask questions ... more broadly'' [P11]} about the impact of ML in society.

Brokers in civil society organisations mostly engaged participants after a system was deployed, in the context of advocacy about (or against) an existing system. Civil society interviewees envisaged projects as opportunities to build participants' confidence to ask questions about technology, or technology policy, in areas that may previously have been obscure: {\itshape "there are a lot of people who may have questions about [technology policy] or be uncertain, but they lack the confidence to ask hard questions to push back." [P3]}

In a few projects conducted in the Global South, brokers engaged with people interacting with deployed systems to surface problems with those systems. While acknowledging that the issue is not limited to ML systems, brokers in these projects noted that the relevant systems were often developed by organisations based in the Global North and deployed in (or on) the local community. In those projects, brokers sought the participation of domain experts (non-data scientists) who were frustrated and reconciling with the systems. As one interviewee noted, {\itshape ``the amount of work that you have to do in order to get these systems to work for you'' [P5]} is an under-recognised form of participation. To capture this information, interviewees suggested using individual stories to clarify the experience of broader collectives: {\itshape ``When we talk about communities impacted by data, for example, that is too abstract. But if we ground it in the experience of a person who belongs to that community, then you understand that community in the abstract better as well.'' [P5]}

\subsection{Infrastructure to Express Interests Through Data}

Once a problem is formulated, data collection is the first opportunity for brokers to engage participants in the model iteration process. However, interviewees who had worked with participants to collect data for an ML artefact or system noted the challenge of getting participants to {\itshape ``express their concerns and needs correctly'' [P15]}, or, at least, as data in a form a machine would understand. To overcome this challenge, several interviewees created infrastructure to support the data collection stage, including tutorials and training: {\itshape ``"If you know how AI works, then it's much clearer why we need your data and also why we need to have the data in a particular way. For example, why we ask [participants] to contribute objects in various locations and in various kind of lighting conditions... if that isn't clear, then the data is rubbish, and then you get a rubbish AI system." [P7]} Others included feedback mechanisms for participants to report whether the collection process was clear and whether data reflected participants' interests (not only ML needs).

Similarly, labelling and annotation with participants challenged brokers to translate contextual information related to labels into machine-readable form. Despite the valuable knowledge participants brought to the annotation process, some brokers reported that the translation challenge limited the scope of engagement. In these cases, interviewees reported that participants became frustrated when conversations narrowed to the data formats required for classification models, instead of addressing the participants' problems and interests: {\itshape ``there are specific ways that you're taught to engage communities and sometimes the community just doesn't want to engage that way.'' [P16]} Additionally, several interviewees highlighted a lack of emphasis from their organisations on their ability to facilitate participatory activities during data labelling, compared to their technical skills in data processing and modelling: {\itshape ``many of the ones who are designing these tasks, do not necessarily feel that they have the right background or training ... that's a set of skills that can take a very long time to develop.'' [P12]}

\subsection{Decoupling Model Development From Model Evaluation}

While brokers noted that data scientists make {\itshape ``a lot of choices.. even if you have a very well-developed set of requirements'' [P11]} across data processing, model training and fine-tuning, most interviewees identified model development as the {\itshape "most challenging stage" [P11]} to involve participants. Some interviewees referred to frameworks facilitating the inclusion of end users and people affected by algorithmic systems during these activities ({\itshape e.g.}, translating high-level design principles into development decisions \cite{Zhu2018-di} and developing interfaces to allow participants to make preferences clear during model development \cite{Lee2019-ke,Cheng2021-gy}). However, interviewees noted there are few dependable tools or methods: {\itshape ``I just don't think the tools are there yet to make [Participatory ML] easy.'' [P10]}

Several interviewees offered two reasons for prioritising participant involvement during model evaluation rather than model development: 1) model development activities are technically complicated, requiring infrastructure and resources to support non-data scientists to participate beyond that of data collection and labelling; 2) model development may be less consequential for the performance and value of resulting systems for participants than earlier stages: {\itshape ``Once you have labelled data, you are most of the way there.'' [P1]} Therefore, some brokers elevated the voices of end users engaged as participants during model evaluation: {\itshape ``Evaluation is really important to have people involved because that's when you're saying, does this actually work? But maybe model development isn't as important ... because I can create iterations quickly and modify them based on people's feedback.'' [P4]} However, perspectives on participant inclusion for model evaluation diverged. Some suggested re-engaging data collectors for a technical assessment of the model's results. In contrast, others---especially in areas with specialist expertise ({\itshape e.g.}, health, law)---argued that evaluation should be performed by the intended users of a system ({\itshape e.g.}, general physicians, not specialist physicians who collected and labelled data), focusing on how an application embedded with a model operates in real-world scenarios.

Within projects conducted in the Global South, some interviewees highlighted a tension between building and deploying {\itshape ``good enough'' [P14]} models and those perfectly tailored for specific contexts. In these projects, model evaluation was a critical opportunity for participants to exercise power over the development process, particularly where data scientists engaged in the project were not based in the local community. By drawing on their domain expertise, participants could {\itshape ``take control of testing and validating outputs'' [P6]} and decide whether a model improved upon existing systems and was ready for deployment. As one broker of such projects noted: {\itshape ``This challenges some of the narratives we have around participatory AI---the need for [the model] to be perfect and ... who makes the decisions about when a model is good enough ... to be offered and used.'' [P14]}

\subsection{Scaling Relationships Post-deployment}
A key feature of all projects described by brokers was the transfer of knowledge between data scientists and participants through brokers. Many interviewees described how participatory projects became a mechanism to bridge knowledge gaps in the broader ML development process---for participants, about how ML systems function and possible applications, and for data scientists, about the social, economic, political, and cultural contexts in which ML systems are or will be deployed.

However, many interviewees described encountering disciplinary divides in their projects. At the start of projects, brokers often faced the challenge of convincing data scientists in their teams to go through the process of building relationships with participants: {\itshape ``there were times where it was really frustrating for [the data scientists] because it involved rethinking the typical pipeline that they follow. And there was less time for them to spend on the technical questions that they really thrive on.'' [P14]}

Once a project was completed, many brokers reported encountering a tension between desires for generalisability in publication venues and the situated nature of their projects. While participatory projects resulted in {\itshape ``more complex designs [and] … more drawn out work'' [P16]}, the features of participatory projects, such as the quantity and diversity of participants, were rarely recognised in traditional evaluation criteria for ML development. As one interviewee noted: {\itshape ``in terms of engineering, [participation] is not traceable.'' [P11]} While noting challenges for recognition of situated projects within research communities, several interviewees discussed strategies to scale project outcomes---in summary, getting the participatory approach right with one group, then replicating and adapting the approach with others while fine-tuning the ML system: {\itshape "Machine learning scales through mass customisation and mass localisation." [P17]}

\section{Discussion}

Participatory ML projects redefine {\itshape ``who gets to be considered a data scientist’’ [P3]}, recognising practical experience and domain expertise as critical to ML development \cite{DIgnazio2020-zn}. Our findings uncovered {\itshape how} participation is achieved across such projects, focusing on the decisions and actions of participation brokers. While prior research suggests that most projects involve consultation to elicit stakeholder preferences, we focused on projects recognised by the research community for their collaborative approach.

Nonetheless, brokers in these projects encountered two sets of tensions that made collaborative engagement with participants difficult: firstly, the conflict between messy contextual information generated through participation and neat, machine-readable forms of information required in ML; and secondly, the inherent tensions within project contexts, characterised by uneven organisational, economic, and decision-making power between brokers, data scientists, and the end users or stakeholders engaged as participants. In this section, we reflect on these tensions and the strategies our interviewees employed to address them, while advocating for an evolution in the role of brokers.

\subsection{`Fitting' Participation Into the Process of ML}

Our interviews demonstrated the challenges brokers face to `fit' participants into the ML development process. While brokers expressed personal motivations to share power with participants, in most cases, brokers and data scientists brought the requisite technical knowledge and resources to undertake Participatory ML projects. As such, brokers exercised power in deciding who became participants, what they participated in, and how their contributions were implemented.

Brokers described a range of strategies across the development process to navigate power dynamics with participants. Initially, some engaged with existing organisations, rather than forming their own groups, to balance organisational power. Others used prototypes of models during problem formulation, to engage participants who may have been previously unfamiliar with ML. Before data collection, several brokers conducted tutorials and training on data requirements for models, ensuring participants grasped the significance of data to ML and could exert influence over that stage. Furthermore, by prioritising participants' perspectives during model evaluation, some brokers empowered participants to influence decisions regarding the deployment readiness of ML systems.

Nonetheless, the process of `fitting’ participants to the ML development process implies an asymmetric division of learning, with brokers holding power in deciding what participants learn about ML while also gaining insights from participants' situated knowledge \cite{Zuboff2019-ni}. For example, several brokers used demos of models to engage participants at the beginning of projects. However, as \citet{Prabhakaran2020-bp} note, using ML tools during problem formulation may foreclose the decision not to use ML in a project. Where there is a power imbalance between participants and brokers or data scientists during problem formulation, one interviewee argued that {\itshape ``there's often a rush to adopt technical solutions to really adaptive and complex problems.'' [P8]}

Further into the development process, brokers were able to open ML systems to feedback from participants during some stages more than others---{\itshape e.g.}, during data collection and labelling. While data collection and labelling may be an opportunity for participants to learn about ML and have an impact on the shape of systems \cite{Patton2020-te,Suresh2022-vv}, the power to decide what categories exist in the world of ML and the correct categories for labelling rested with brokers \cite{Bowker2000-ot}. Additionally, as noted by several brokers, the bounds of engagement with participants during data collection and labelling were constrained by the instrumental needs of the broader ML development project. The fundamental challenge of balancing the instrumental requirements of ML development while affording participants power over the process indicates the need for evolution in the role of brokers.

\subsection{Re-Imagining Brokers as Activists}

If Participatory ML projects are to change more than the definition of a data scientist and address the democratic deficit in AI \cite{Kulynych2020-et}, then the role of brokers must expand beyond acclimatising intended users to ML. In recent decades, researchers and designers in AR and PD projects have been urged to embrace activism, committing to social change and ``taking a stance'' in support of participants \cite{Bannon2018-ce,Bodker2018-ue,Chatterton2007-rh}. Similarly, \citet{DIgnazio2020-zn} recently outlined an approach to data science guided by activism, which has guided recent work in Participatory ML [{\itshape e.g.}, \citealp{Suresh2022-vv}].

Our findings reveal variation in how brokers engage with participants in projects. While some interviewees focused on gathering information from participants, others engaged in a reciprocal exchange, educating participants about ML and using insights from projects to advocate for participants' interests in broader computing research and development contexts. Such actions indicate the potential for brokers to evolve into activists, forging ongoing relationships with participants to envision "possibilities and alternatives" through ML \cite{Bannon2018-ce}. Below, we briefly outline strategies to guide brokers in embracing this activist role.

\subsubsection{Education and Advocacy for End Users}

Brokers working with end users may scaffold participation through tools to educate and engage end users about ML and its potential applications \cite{Theodorou2021-sx}. We see this as an extension of the educational work several brokers reported conducting in relation to data collection and labelling activities. For example, this may involve engaging intended or potential end users in speculative design activities to contest ML design assumptions [{\itshape e.g.}, \citealp{Alfrink2022-xw,Alfrink2023-jk}]. Additionally, brokers must be attuned to the possibility of ``reverse tutelage'' \cite{Mohamed2020-ol}---not only acquiring information from end users, but also learning about the situations in which ML systems are or will be used, and advocating for the interests of end users in computing research and development contexts.

\subsubsection{From Stakeholder Frustration to Feedback}

Engaging with indirect stakeholders was particularly challenging for the projects included in our study. While many brokers described working with supportive partners, interviewees from civil society and some brokers of Global South projects engaged with frustrated stakeholders to expose points of contention with deployed systems. As \citet{Sambasivan2022-uo} note, domain experts are often reduced to data collectors in AI research, particularly in Global South settings. Elevating such expertise requires attending to frustration, refusal, and dissent concerning ML systems as viable forms of agency and key sources of feedback \cite{Dobbe2021-wq,Zong2023-va}.

To bridge the gap between frustration and feedback first requires brokers to educate indirect stakeholders about ML, similarly to end users, as participating with respect to deployed systems {\itshape ``requires a certain kind of literacy … so that you can actually navigate your data representation.'' [P5]} Additionally, bridging this gap requires the development of methodologies that enable people interacting with ML systems to express their interests in ways that can be input into ML development and related policy processes. For example, this may require further work on transforming qualitative, personal stories into feedback for technical development processes \cite{Singh2022-fa}, and further development of collaborative algorithmic audit platforms to encourage relationships between data scientists and dissenters \cite{Deng2023-bs}.

\section{Conclusion}
In our interview study with 18 participation brokers across a range of organisational settings and locations, we explored how participation is achieved across the stages of a participatory project within the ML design and development context. While we hope our findings will guide future participation brokers, we acknowledge the inherent tension between messy contextual information generated through participation and the rigid data formats of ML workflows, as well as uneven power dynamics within project contexts. Recognising these challenges, we advocate for a more pronounced activist role for participation brokers. This role involves not only `fitting' participants into established engineering processes, but also forging relationships with participants over time to empower participants to envision alternatives and possibilities through ML. By educating and advocating for end users, and attending to frustration and dissent from indirect stakeholders, participation brokers can help steer Participatory ML towards its democratic potential.







\begin{acks}
This research is supported by an Australian Government Research Training Program (RTP) Scholarship and a Florence Violet McKenzie scholarship. Our sincere thanks to the interviewees who shared their knowledge and experience with us: Aleks Berditchevskaia, Anna Brown, Aaron Halfaker, Ken Holstein, Michael Katell, Jennifer Lee, Vukosi Marivate, Rebeca Moreno Jiménez, Rajendran Narayanan, Harini Suresh, Haiyi Zhu, and others who wish to remain anonymous. We also thank our reviewers for their feedback and suggestions.
\end{acks}

\bibliographystyle{ACM-Reference-Format}
\bibliography{main.bib}










\end{document}